\documentclass[twocolumn,showpacs,showkeys,preprintnumbers,amsmath,amssymb,prb]{revtex4}

\usepackage{graphics,color}

\usepackage{graphicx}

\begin{document}

\title{Non-collinear Antiferromagnets and the Anomalous Hall Effect}
\author{J\"urgen K\"ubler$^1$}
 \email{juergen.kuebler@gmail.com}
\author{Claudia Felser$^2$}
\affiliation{%
 $^1$Technische Universit{\"a}t Darmstadt, Germany }%
\affiliation{%
 $^2$Max Planck Institute for Chemical Physics of Solids, Dresden, Germany }%
\date{\today}
\begin{abstract}{The anomalous Hall effect is investigated theoretically for the non-collinear antiferromagnetic order of the hexagonal compounds Mn$_3$Ge and Mn$_3$Sn using various planar triangular magnetic configurations as well as  unexpected non-planar configurations. The former give rise to  anomalous Hall conductivities (AHC) that are  found to be extremely anisotropic. For the planar cases the AHC is connected with  Weyl-points in the energy-band structure, which are described in detail. If this case were observable in Mn$_3$Ge,  a large AHC of about $\sigma_{zx}\approx 900~ (\Omega \rm cm)^{-1}$ should be expected. However, in Mn$_3$Ge it is the non-planar configuration that is energetically favored, in which case it gives rise to an AHC of $\sigma_{xy}\approx 100~ (\Omega \rm cm)^{-1}$}. The non-planar configuration allows a quantitative evaluation of the topological Hall effect (THE) that is seen to determine  this value of $\sigma_{xy}$  to a large extent. For Mn$_3$Sn  it is  the planar configurations that  are predicted to be observable. In this case the AHC can be as large as $\sigma_{yz}\approx250 ~(\Omega \rm cm)^{-1}$.    

\end{abstract}
\pacs{75.50.Ee, 75.47.Np, 73.22.Gk, 75.70.Tj}
\keywords{Antiferromagnets AHE}

\maketitle

\textbf{\textit{Introduction}}

The well-known Hall effect \cite{ashcroft} is observed in all conducting materials, but is especially large in ferromagnets, where it is dominated by a contribution that is not due to the Lorentz force. The latter is dissipationless and is called the anomalous Hall effect (AHE). This effect was explained long ago by Karplus and Luttinger, \cite{karplus} who invoked spin-orbit coupling and perturbation by the applied electric field to expose an additional term to be added to the usual electron velocity. Rather recently this additional term was discovered to be related to the Berry curvature in momentum space. \cite{fang1}  It is an important correction to all transport properties \cite{xiao} that rely on the velocity. In particular it describes the leading contribution to the AHE. \cite{nagaosa}

Usually the AHE in a ferromagnet is assumed to be proportional to the magnetization although this cannot be taken too literally. In fact, Chen \textit{et al.} \cite{chen} very recently found theoretically that the AHE should be observable in certain non-collinear antiferromagnets with zero net magnetization, provided that some symmetries are absent. They predicted this effect for the cubic antiferromagnet Mn$_3$Ir which was calculated to have a rather large anomaölous Hall conductivity. The non-collinear antiferromagnetism in Mn$_3$Ir is of the same kind as that described for Mn$_3$Sn some time ago.\cite{kubler88} The prediction by Chen \textit{et al.} \cite{chen} for Mn$_3$Ir can be extended to the family of cubic, non-collinear antiferromagnets Mn$_3$Sn, Mn$_3$Pt and Mn$_3$Rh for which the calculations of the type used in ref.\cite{chen} have been repeated; they support their findings. 


\begin{figure}
\includegraphics[scale=0.3]{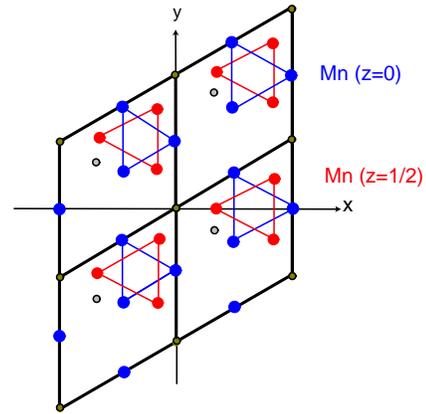}
\caption{\label{fig1}Projection on the basal plane of the hexagonal crystal structure of  Mn$_3$Z (Z=Ga, Sn, and Ge). The small circles represent Z in the basal plane ($z=0$) and in a plane at $z=0.5c/a$. The large colored spheres represent Mn in the same two planes. The triangles are drawn to guide the eye. The magnetic ordering is not shown at this stage.}
\end{figure}

Recent experimental work on materials that show a large exchange bias was conducted for the hexagonal compound Mn$_3$Ge.\cite{qian} This is one of a family of non-collinear antiferromagnets that are related to Heusler compounds. It supplies a wealth of different non-collinear magnetic configurations and, by means of small deviations in the stoichiometry, is close in the phase diagram   to tetragonal ferrimagnets, thus making it an interesting case for applications.

 The system Mn$_3$Z (Z=Ga, Sn, and Ge) can be viewed as Heusler compounds which occur in different structures. A hexagonal phase having the symmetry $P6_{3}/mmc$ (space group number 194) has been grown quite some time ago by annealing the crystals at high temperatures.  \cite{kren70,kadar71,gupta70,yamada}
Tetragonal phases were obtained for Mn$_3$Ga and Mn$_3$Ge by annealing at low temperatures. \cite{kren70,kadar71,Baalke} These different phases have markedly different magnetic properties, the tetragonal ones being ferrimagnetic while the hexagonal crystals are antiferromagnetic with a very small ferromagnetic component. The latter form a kagome lattice with a triangular coupling. The triangular arrangement is sketched in Fig. \ref{fig1} omitting at this stage the directions of the magnetic moments. 

\begin{figure}[h]
\includegraphics[width=0.7\linewidth]{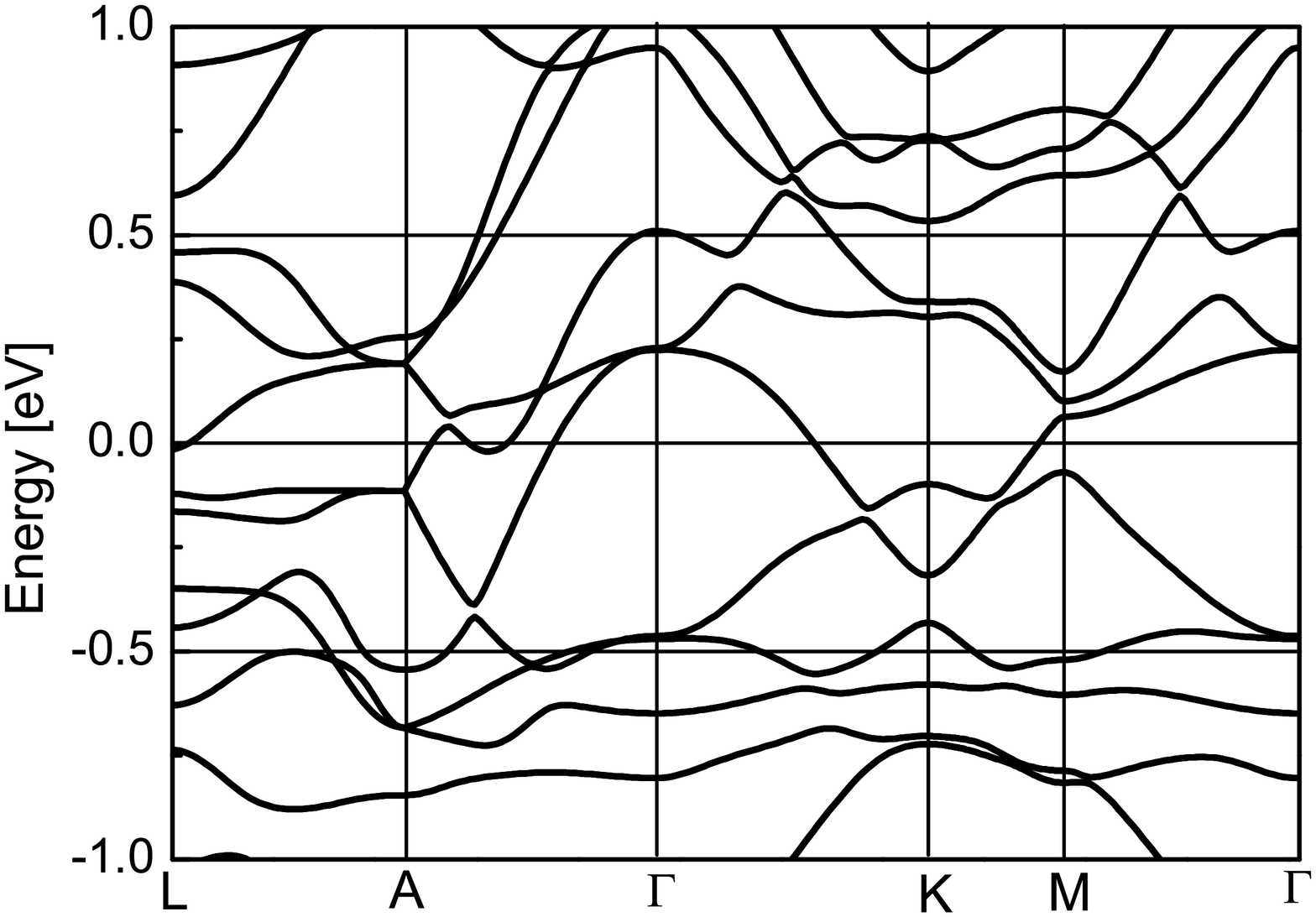}\includegraphics[width=0.3\linewidth]{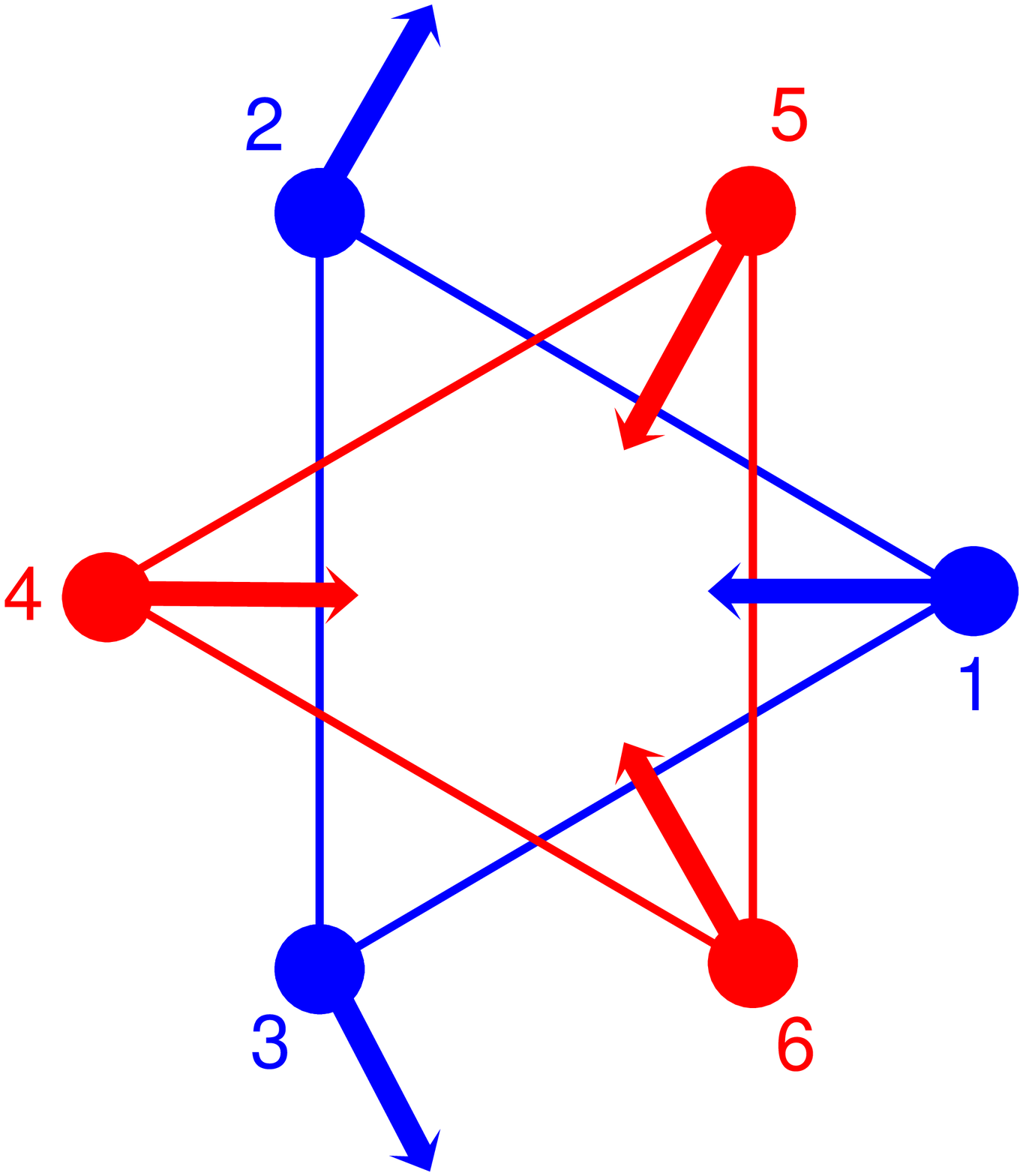}
\caption{ \label{fig2} Band structure of Mn$_3$Ge for the antiferromagnetic structure whose magnetic configuration is sketched on the right. The AHE vanishes for this case. The band structure of Mn$_3$Sn for this configuration is very similar to that of Mn$_3$Ge.}
\end{figure}

The various magnetic and structural properties of Mn$_3$Z (Z= Ga, Sn, Ge) were recently reviewed  theoretically by Zhang \textit{et al}. \cite{zhang} In the present study we focus our attention on the hexagonal phase of Mn$_3$Ge and Mn$_3$Sn.

\textbf{\textit{Magnetic Properties of hexagonal \textrm{Mn$_3$Ge} and
\textrm{Mn$_3$Sn}  and the AHE}}

The early experimental work \cite{kren70,kadar71,gupta70,yamada} and later theoretical studies \cite{sticht,sandr96,zhang} illuminated the unusual and interesting magnetic structure of Mn$_3$Ge and Mn$_3$Sn. There are various possible triangular configuration when the magnetic moments point in the basal plane shown in Fig.\ref{fig1}. Because the Mn-Mn bonds between neighboring layers are somewhat shorter than the in-plane bonds, the interlayer magnetic coupling is important. Various configurations are shown in the recent paper by Zhang \textit{et al.} \cite{zhang} and also in older work by Sticht \cite{sticht} as well is by Sandratskii and this author. \cite{sandr96} Most of these configuration are degenerate as long as spin-orbit coupling (SOC) is ignored. This changes markedly when SOC is included in the calculations, a fact that was convincingly explained in ref. \cite{sandr96}

We add to the previous work a determination of the Hall conductivity due to the AHE by computing the Berry curvature in momentum space. This is a vector obtained from the curl of the Berry connection given by 
\begin{equation}\label{eq1}
\mathcal{A}(\mathbf{k})=i\sum_n \langle{u_{n{\bf k}}}|\nabla_k|{u_{n{\bf k}}}\rangle ,
\end{equation}
where $u_{{n\bf k}}(\mathbf{r})$ is the crystal-periodic
eigenfunction having wave vector \textbf{k} and band index $n$.
The sum extends over the occupied states which for metals vary
with \textbf{k} since there is a Fermi surface. The Berry curvature is obtained from
\begin{equation}\label{eq2}
\bf{\Omega}(\mathbf{k})=\nabla_{\mathbf{k}}\times\mathcal{A}(\mathbf{k}).
\end{equation}
The numerical evaluation is done by  using the wave functions from density functional calculations \cite{williams,kub2009}. This part of the calculations is done as in ref.\cite{kubler12}. The Hall conductivity follows from the Berry curvature as 
\begin{equation}\label{eq3}
\sigma_{\ell m}=\frac{e^2}{\hbar}
\,\int\frac{d\mathbf{k}}{(2\pi)^d}
\Omega_o(\mathbf{k})f(\mathbf{k}),
\end{equation}
where $f(\mathbf{k})$ is the Fermi distribution function,
$\Omega_o(\mathbf{k})$ is the $o$-component of the Berry curvature
for the wave-vector \textbf{k} and the components $\ell, m ,o$ are to be chosen cyclic.\cite{xiao}

\textbf{\textit{Results for the anomalous Hall conductivity}}

 First, some general remarks on the Hall conductivity  are in order. The conductivity vanishes if spin-orbit coupling is ignored in the calculations, just as stated in ref.\cite{chen}~ Furthermore, the Hall conductivity is found to be remarkably anisotropic for all configurations, for which the magnetic moments lie in the hexagonal plane. Thus, in particular, the Berry curvature vector in the $z$-direction vanishes. If the conductivity is to be measured, this will require a somewhat unusual choice of the Hall cross, \textit{i.e.} the electric field must be in the $z$-direction.
For Mn$_3$Ge, however, we find a prominent non-planar configuration with a sizable value of $\sigma_{xy}$. This configuration has not been anticipated before and is discussed below.

\begin{figure}[h]
\includegraphics[width=0.3\linewidth]{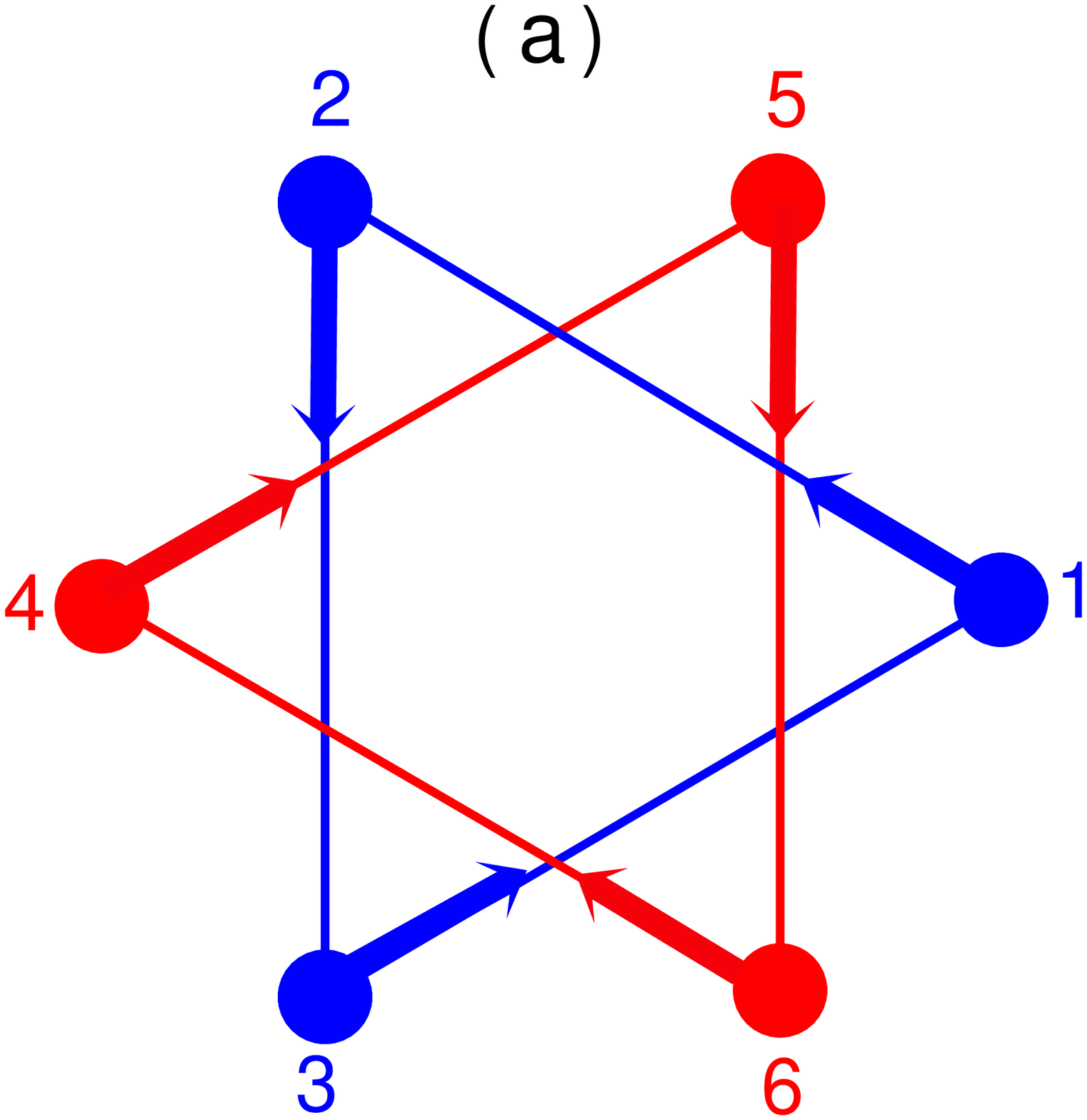}\includegraphics[width=0.32\linewidth]{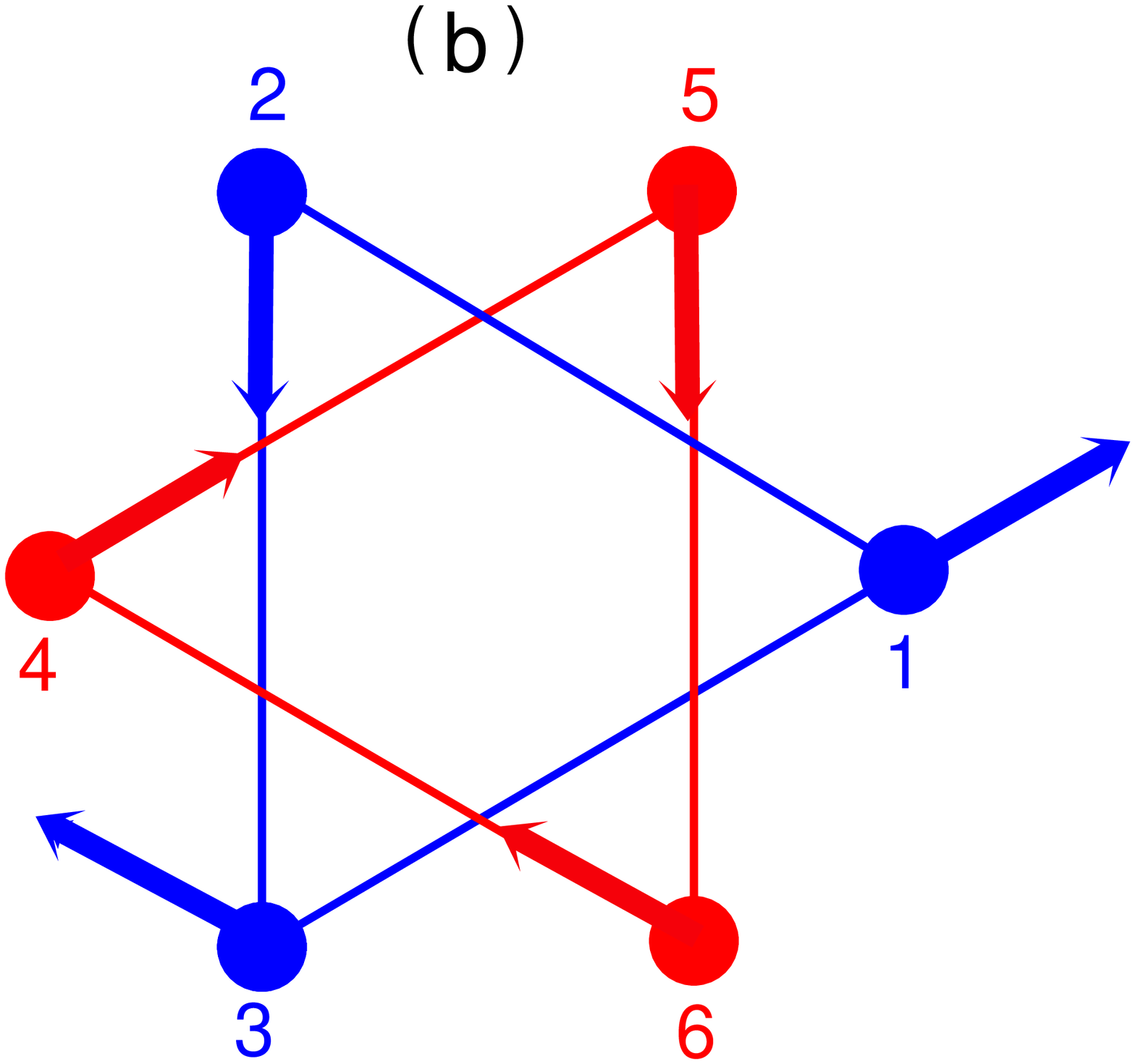}
\caption{ \label{fig3} Two triangular magnetic configurations. (a) Demonstrates opposite winding (or chirality) of the magnetic moments in the two triangles. (b) The configuration of (a) after a self-consistent calculation. Note the change of the directions of the moments of atom 1 and 3, which leads to the chirality of the two triangles to become the same. }
\end{figure}

\begin{table}
\caption{\label{t1} 
The total energy of the listed magnetic configuration, $\Delta E$, the magnetic moments of Mn, $M_{\rm Mn}$, in $\mu_{\rm B}$, estimates of the ferromagnetic spin and orbital moments, $M^{sp}$, $M^{orb}$  in $\mu_{\rm B}$, and the Hall conductivities, $\sigma_{yz}$, $\sigma_{zx}$ and $\sigma_{xy}$ in $(\Omega \rm cm)^{-1}$. THE gives the contribution from the topological Hall effect. The lattice constants were taken from experiment \cite{kadar71,zhang}: for Mn$_3$Ge $a=0.536 $ nm, $c/a=0.80598$ and for Mn$_3$Sn $a=0.5665$ nm, $c/a=0.79982$.} 
\begin{center}
\begin{tabular}{lccccccc}
\hline \
Config. & $\Delta E$[meV]&  $M_{\rm Mn}$ & $M^{sp}$  &$M^{orb}$ & $\sigma_{yz}$ &$\sigma_{zx}$&$\sigma_{xy}$\\
\hline 
\textbf{Mn$_3$Ge}& & & & & & & \\ 
Fig. \ref{fig2}    & 744 &2.258 &0     &0    & 0  &0   &0  \\
Fig. \ref{fig3}(b) & 1   &2.738 & 0.007&0.03  &667 &379 &0 \\
Fig. \ref{fig5}(c) & 0   &2.738 & 0.004&0.03 &607 &1   &0 \\
Fig. \ref{fig7}    &-27  &2.738 & 0.002&0.02 &231 &-965&104 \\
THE                &     &      &      &     &  6  & 4   & 85   \\  
\hline
\textbf{Mn$_3$Sn}& & & & & & & \\ 
Fig. \ref{fig3}(b) & 0   &3.121 & 0.003&0.001  &248 &129 &0 \\
Fig. \ref{fig5}(c) & 1   &3.120 & 0    &0.001 &256 &17   &0 \\
Fig. \ref{fig7}    & 2   &3.121 & 0.01 &0.001 &95  &111  &8 \\
THE                &     &       &      &      &0.1 &0.1&4
 \\
\hline \hline

\end{tabular}
\end{center}
\end{table}

Next, in Fig.~\ref{fig2} we add the band structure to our discussion. The configuration shown is self-consistent and acquires no ferromagnetic component. The band structure is doubly degenerate, except for some  symmetry points where it may be higher. This and the fact that the magnetic moments sketched in Fig.~\ref{fig2} are pairwise antiparallel indicates that this is the case of a bipartite lattice where time-reversal symmetry is not broken. The AHE vanishes by symmetry. This is the normal case for antiferromagnets as was also discussed by Chen \textit{et al.}\cite{chen}  Since the total energy of this non-collinear spin configuration is rather high (see Table I), we turn to more interesting cases.

\begin{figure}[h]
\includegraphics[width=1.0\linewidth]{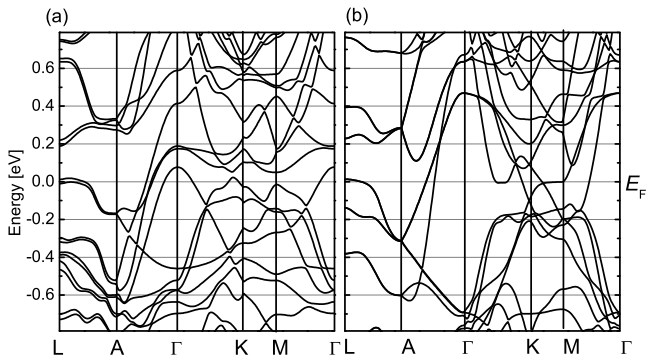}
\caption{ \label{fig4} (a) Band structure of Mn$_3$Ge of the opposite-chirality case given in part (a) of Fig.~\ref{fig3}. (b) Band structure of the equal-chirality case given in part (b) of Fig.~\ref{fig3}. Note especially the bands in the range K to M.  }
\end{figure}

\begin{figure}[h]
\includegraphics[width=0.7\linewidth]{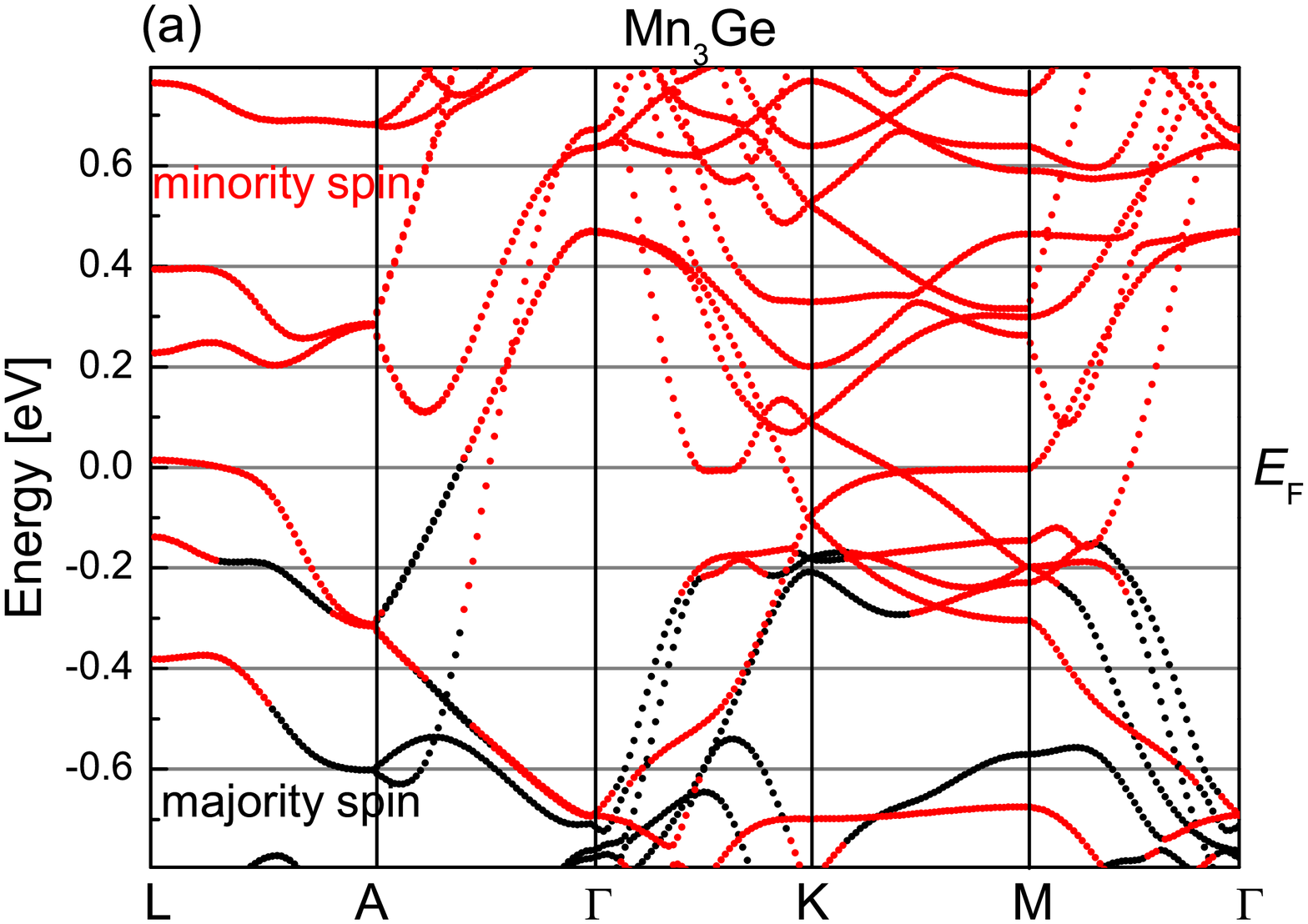}
\includegraphics[width=0.7\linewidth]{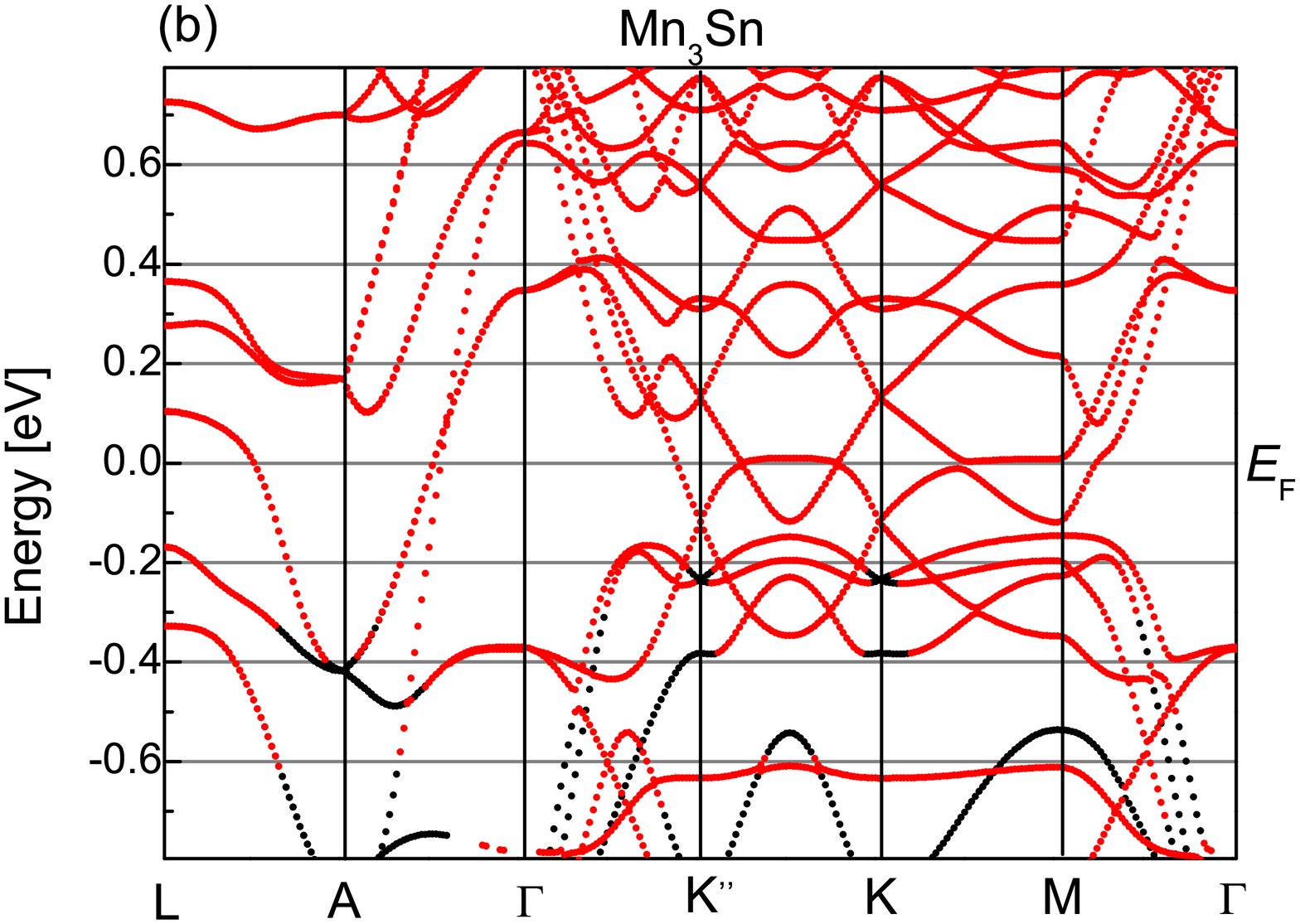}
\includegraphics[width=0.3\linewidth]{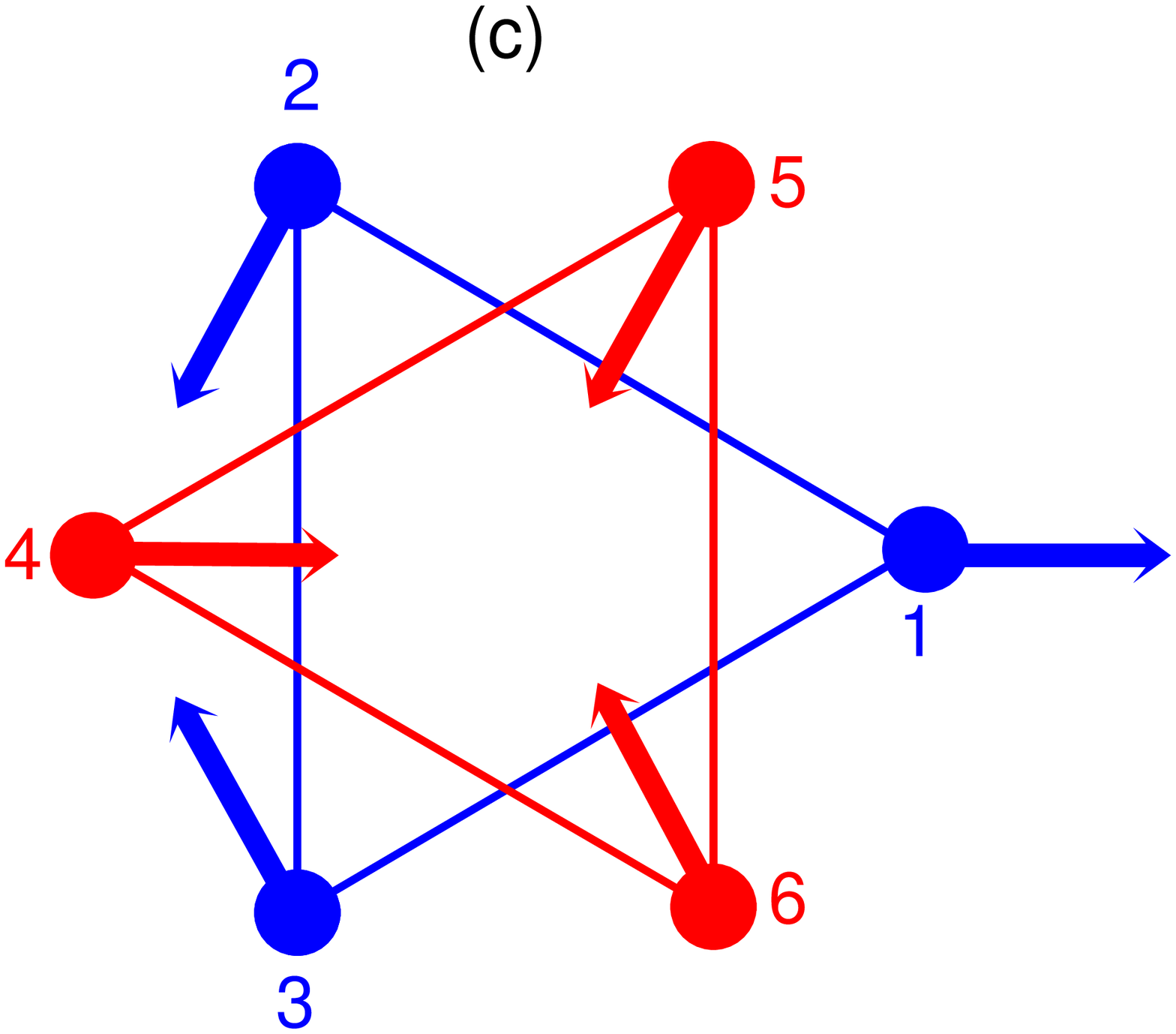}
\caption{\label{fig5} (a) Spin-resolved band structure of Mn$_3$Ge for the magnetic configuration shown in panel (c) along the standard symmetry lines. (b) Spin-resolved band structure of Mn$_3$Sn for the magnetic configuration shown in (c) along standard symmetry lines extended by the section $\Gamma$- $K^{''}$. See Fig.\ref{fig6}(b) for the definition of the label $K^{''}$.}
\end{figure}

\begin{figure}[h]
\includegraphics[width=0.75\linewidth]{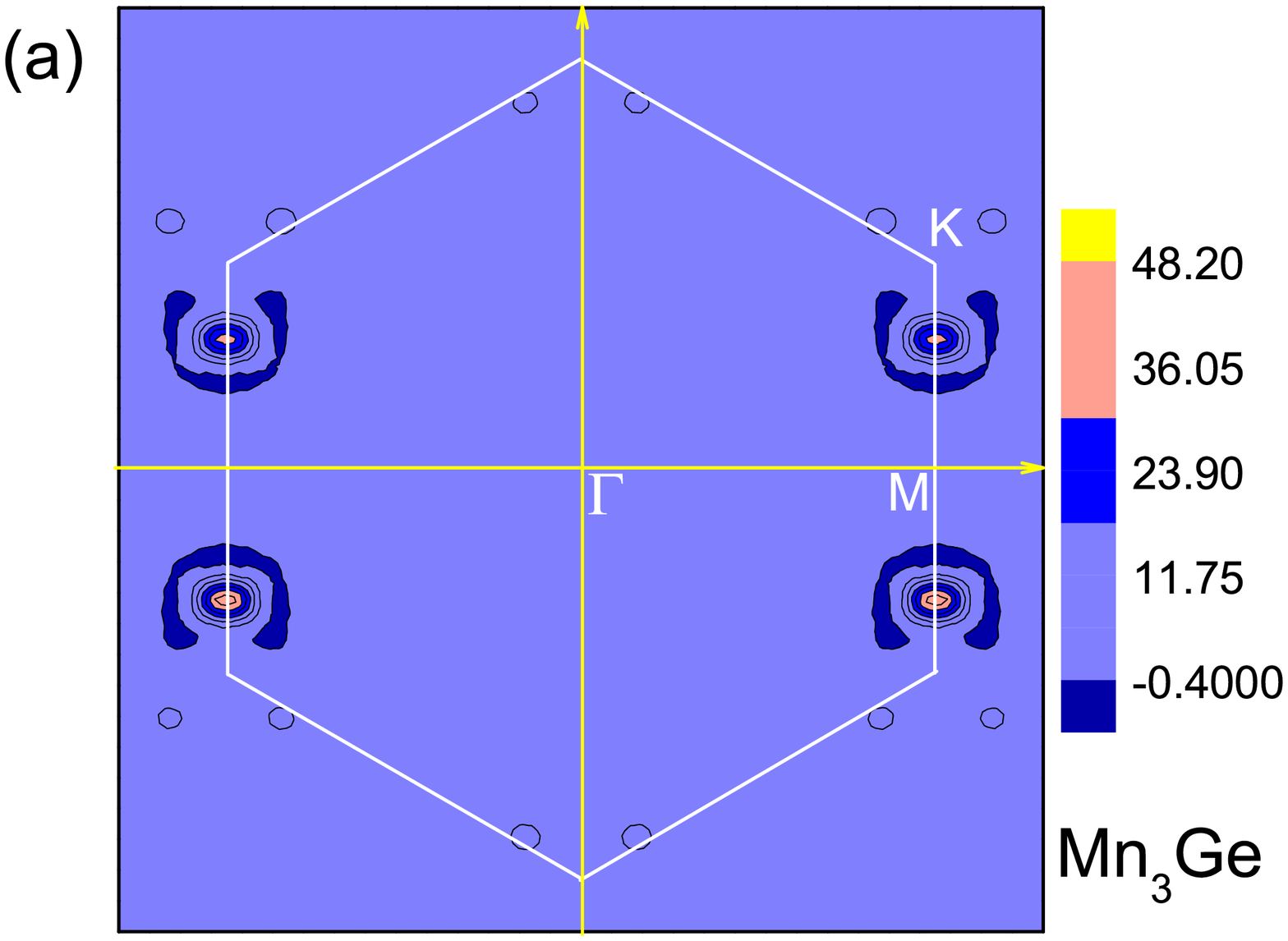}
\includegraphics[width=0.78\linewidth]{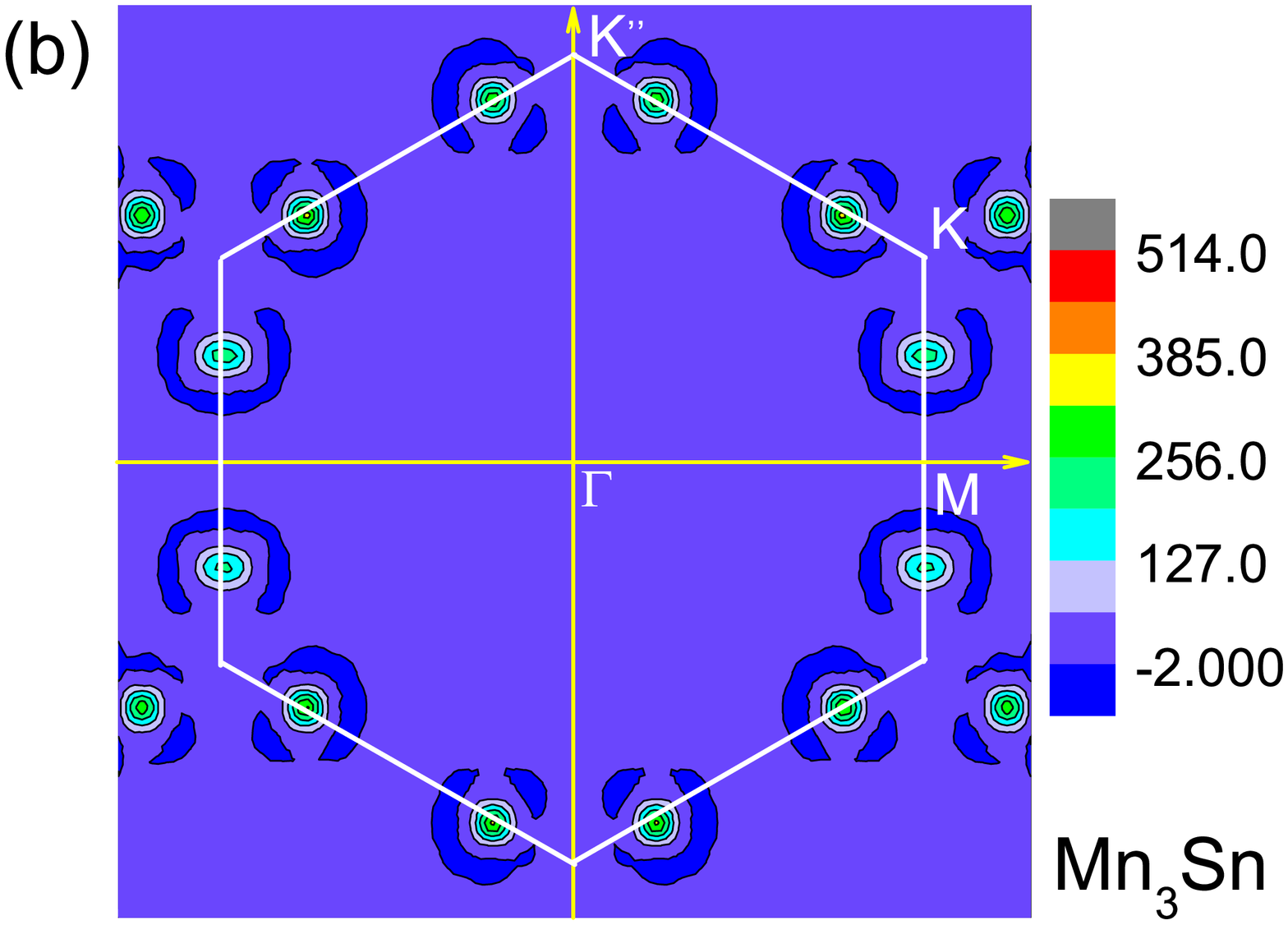}
\caption{\label{fig6} (a) The Berry curvature divided by $2\pi$  of Mn$_3$Ge calculated in the $k_z=0$ - plane for the magnetic configuration shown in panel Fig.~\ref{fig5}(c). (b) same as (a) for Mn$_3$Sn .}
\end{figure}

\begin{figure}[h]
\includegraphics[width=0.6\linewidth]{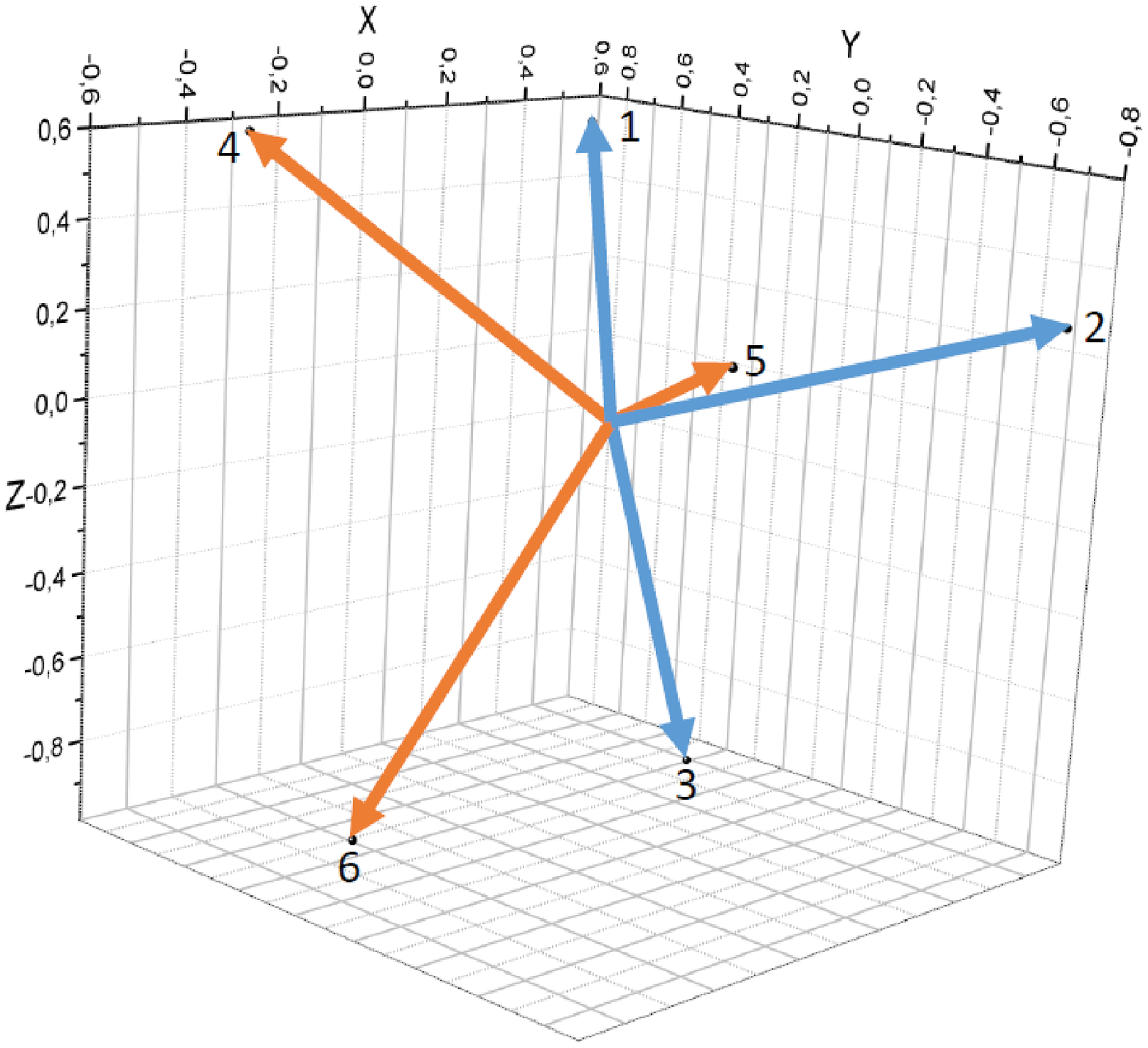}
\caption{\label{fig7} Non-planar configuration of the magnetic moments. For technical reasons the six direction vectors are shifted such that they originate at the origin of the reference system.   }
\end{figure}

    We now focuss our attention on a general property of a non-collinear magnetic configuration, \textit{i.e.} on the winding or chirality. A special choice is shown in Fig.~\ref{fig3} part (a) where the opposite chirality of the two groups of the magnetic moments is easily seen. Although quite seductive the so chosen configuration (a) is not stable, which means it "unfolds" in the self-consistent calculation to the configuration depicted in part (b). This is due to the interlayer exchange interactions that rotate two of the magnetic moments. Both parts of the configuration (b) now have the same chirality. The conductivities are finite in each case, being for the non-selfconsistent case (a) of the order of 100 $(\Omega \rm cm)^{-1}$ but much larger for (b) as seen in Table \ref{t1}. On first sight one tends to attribute this large change  to the appearance of  a very small ferromagnetic component that develops in the self-consistent calculation (see Table \ref{t1}). However, a different reason emerges from a special property of the  band structure, which is graphed in Fig.~\ref{fig4} for the two cases (a) and (b) of Fig.~\ref{fig3}. Drawing the attention to the band structure in the range K to M, we see there is nothing unusual in case (a), but in case (b) two non-degenerate bands cross at the Fermi edge. This crossing point has recently been named Weyl point \cite{wan} which is expected to result in interesting topological properties, especially those concerning the anomalous Hall effect. 

To elucidate the role of the Weyl point we show in part (a) of  Fig.~\ref{fig5} the spin-filtered band structure of Mn$_3$Ge and in part (b) that of Mn$_3$Sn. In both cases the self-consistent configuration depicted in  Fig.~\ref{fig5} (c) was used. For Mn$_3$Ge the  band crossing occurs as in Fig.~\ref{fig4}(b) between the points $K$ and $M$ (not drawn to scale). For Mn$_3$Sn, however,  the bands are slightly gapped between $K$ and $M$, instead the band crossings here occur twice between the points $K^{''}$ and $K$. In both cases the bands at the Fermi energy are non-degenerate minority-spin electron bands.

To round up the physical picture we calculated  the Berry curvatures in the  plane that is extended enough to show the entire symmetry of the hexagonal basal plane of the Brillouin zone(BZ). The results are graphed in Fig. \ref{fig6} (a) for Mn$_3$Ge and in \ref{fig6} (b) for Mn$_3$Sn. The Weyl points appear distinctly at the places corresponding to the crossing points in the band structure. Closer inspection shows for Mn$_3$Sn that the spots originating from the crossings between $K^{''}$ and $K$ are much more pronounced than the spots seen between $K$ and $M$, where the states are slightly gapped. The role of the Weyl points for the AHE is interesting  and has been discussed controversially in the recent literature by Haldane \cite{hald}, Chen \textit{et al.} \cite{chen1}, and Vanderbilt \textit{et al.} \cite{vander} 

Adding to the discussion of chirality in connection with Fig.~\ref{fig3} we observe that the chirality of the two triangles shown in Fig.~\ref{fig5}(c)
is the same. This can be changed by interchanging the direction of arrows 2 and 3. After selfconsisting this configuration the AHC is found to vanish. The Weyl points, however, remain visible in the berry curvature, but the phase relations change such that the contributions to the berry curvature cancel out in the Brillouin zone.  

The discussion of the AHE is not yet complete. By trial and error we discovered  a non-planar antiferromagnetic  configuration that gives rise to a large AHC. For Mn$_3$Ge this configuration is illustrated in Fig.\ref{fig7}, for Mn$_3$Sn it differs somewhat. For Mn$_3$Ge the total energy favors this non-collinear structure, which has not been discussed previously. Since the anisotropy is much less pronounced (see Table \ref{t1}) it is plausible that its AHE is easier to observe, whereas for Mn$_3$Sn  it will be the planar cases  that should be measurable. We could not find a Weyl point to be connected with the non-planar structures.

In connection with the non-planar magnetic configuration the contribution to the AHC arising from the topology, the so-called topological Hall effect (THE) \cite{onoda} can be obtained. This is done by calculating the Hall conductivity omitting spin-orbit coupling. The results are included in Table \ref{t1}. For Mn$_3$Ge the value for $\sigma_{xy}$ is quite significant and is seen to be the dominant contribution to the conductivity.  Its being much larger than that for Mn$_3$Sn is explained by the difference  in the 'spin chiralities', which for non-planar configurations may be defined by \cite{onoda} $\kappa={\bf n}_i\cdot({\bf n}_j\times{\bf n}_k)$, where ${\bf n}_i$ is the unit vector giving the direction of the moment at site $i$ and $(i,j,k)$ is chosen to be (1,2,3) or (3,4,5). In the case of Mn$_3$Ge the configuration shown in Fig. \ref{fig7} gives $\kappa \approx 0.88$, whereas for Mn$_3$Sn we obtain only $\kappa \approx 0.04$. For the planar magnetic configurations $\kappa=0$.

Very recently exciting experimental work on the AHE in non-collinear antiferromagnetic Mn$_5$Si$_3$ \cite{surgers}  came to our attention.
Since the crystal structure of Mn$_5$Si$_3$ is considerably more complicated than that of Mn$_3$Sn or Mn$_3$Ge it appears that experimental work on the latter compounds could be quite rewarding, too.

\textbf{\textit{Summary}}
The AHE in the non-collinear antiferromagnetic compounds Mn$_3$Ge and Mn$_3$Sn is described in detail. Two different sets of magnetic configurations are relevant: triangular planar and
 non-planar. For Mn$_3$Sn the total energy favors a planar structure for which the AHC is predicted to be markedly anisotropic. For Mn$_3$Ge the non-planar case dominates, for which the topological Hall effect contributes significantly to the conductivity $\sigma_{xy}$.  
\acknowledgments
Useful comments by Binghai Yan and an inspiring discussion with Stuart Parkin are gratefully acknowledged.

\end{document}